\documentclass[twocolumn,twoside,slac_two]{revtex4}
\usepackage{graphicx}
\usepackage{fancyhdr}
\begin{document}

%Title of paper
\title{Southern-Hemisphere AGN Monitoring on (Sub-)Parsec Scales:\\The TANAMI Program}

\author{C.~M\"uller}
\affiliation{Erlangen Centre for Astroparticle Physics, Erwin-Rommel Str. 1, 91058 Erlangen, Germany}
\author{M.~Kadler}
\affiliation{Institut f\"ur Theoretische Physik und Astrophysik, Universit\"at W\"urzburg, 97074 W\"urzburg, Germany}
\affiliation{CRESST/NASA Goddard Space Flight Center, Greenbelt, MD 20771, USA}
\author{R.~Ojha}
\affiliation{Astrophysics Science Division, NASA Goddard Space Flight Center, Greenbelt, MD 20771, USA}
\author{J.~Blanchard}
\affiliation{University of Tasmania, Hobart 7001, Australia}
\author{M.~B\"ock}
\affiliation{Erlangen Centre for Astroparticle Physics, Erwin-Rommel Str. 1, 91058 Erlangen, Germany}
\author{M.~Dutka}
\affiliation{The Catholic University of America, 620 Michigan Ave., N.E.,  Washington, DC 20064, USA}

% %
\author{E.~Ros}
\affiliation{Dept. d'Astronomia i Astrof\'{\i}sica, Universitat de Val\`encia, E-46100 Burjassot, Val\`encia, Spain}
\affiliation{Max-Planck-Institut f\"ur Radioastronomie, Auf dem H\"ugel 69, 53121 Bonn, Germany}
\author{J.~Wilms}
\affiliation{Erlangen Centre for Astroparticle Physics, Erwin-Rommel Str. 1, 91058 Erlangen, Germany}
\author{and the TANAMI collaboration}
\affiliation{}

%%%%%%%%%%%%%%%%%%%%%%%%%%%%%%%%%%%%%%%%%%%%%%%%%%%%%%%%%%%%%%%%%%%%%
\begin{abstract}
The Very Long Baseline Interferometry (VLBI) monitoring program TANAMI provides bi-monthly,
dual-frequency (8\,GHz and 22\,GHz) observations of extragalactic jets with milliarcsecond resolution south of $-30^\circ$ declination using the Australian Long Baseline Array (LBA) and
additional radio telescopes in Antarctica, Chile, New Zealand and South
Africa \citep{Ojha2010a}. 
Supporting programs provide multiwavelength coverage of the 
\textit{Fermi}/LAT sources of the TANAMI sample, in order to construct simultaneous
broadband spectral energy distributions (SEDs), as well as rapid follow-ups of high energy flares.
The main purpose of this project is to study the radio-gamma-ray connection seen in the jets of active galactic nuclei (AGN) via simultaneous monitoring of their VLBI structure and broadband emission in order to distinguish between different proposed emission models.

Here we give a brief description of the TANAMI program and will then focus on its current status: 
(1) We present some results on the first simultaneous dual-frequency images of the whole sample resulting
in spectral index maps of the parsec-scale core-jet structure.
(2) The TANAMI array allows us to observe the closest radio galaxy Centaurus~A with unprecedented 
high angular resolution resulting in the best-ever image of an AGN jet. We constructed the best resolved spectral index map of its jet-counterjet
system revealing multiple possible production sites of $\gamma$-rays recently detected by \textit{Fermi}/LAT. With the first epochs of the TANAMI monitoring, we can study the proper jet motion of individual jet components of Cen~A on sub-parsec scales.
(3) Since the launch of \textit{Fermi}/LAT we added newly detected $\gamma$-ray bright AGN to the TANAMI observing list which is built as a combined radio and  $\gamma$-ray selected sample. For most of these sources the TANAMI observations obtain the first VLBI images ever made.  
\end{abstract}
%%%%%%%%%%%%%%%%%%%%%%%%%%%%%%%%%%%%%%%%%%%%%%%%%%%%%%%%%%%%%%%%%%%%%
\maketitle
\thispagestyle{fancy}
%
%%%%%%%%%%%%%%%%%%%%%%%%%%%%%%%%%%%%%%%%%%%%%%%%%%%%%%%%%%%%%%%%%%%%%
\section{Project Outline}\label{sec1}
Blazars are a radio-loud subset of Active Galactic Nuclei (AGN) being highly variable across the whole electromagnetic spectrum. VLBI observations show that they are very compact objects on parsec-scales with powerful, highly relativistic outflows, called jets, which are oriented close to our line of sight. Multiwavelength observations are crucial to answer fundamental questions about the formation, composition and emission mechanism of these extragalactic jets. Studies of the $\gamma$-ray bright blazar population suggested a close connection between the radio and the high energy emission, spectral changes, outbursts and the ejection of parsec-scale jet components \citep{Dondi1995}. Different emission models are proposed to explain the observed $\gamma$-ray emission of (blazar) jets (see \cite{Bottcher2007} for a review).
Simultaneous broadband spectral energy distribution (SED) measurements across the electromagnetic spectrum are required to discriminate between these models. As the only method which can spatially resolve the parsec scale jet structure and separate them from other AGN parts, VLBI monitoring of blazars completes these investigations by providing direct measurements of jet properties like component ejection times, apparent speed, or opening angle. 

The \textbf{T}racking \textbf{A}ctive Galactic \textbf{N}uclei with \textbf{A}ustral \textbf{M}illi\-arcsecond \textbf{I}nterferometry program\footnote{\url{http://pulsar.sternwarte.uni-erlangen.de/tanami}} provides quasi-simultaneous, dual-frequency VLBI observations of extragalactic jets south of $-30^\circ$ declination \citep{Ojha2010a}.
The unique Southern Hemisphere TANAMI array is composed of the Long Baseline Array in Australia and additional telescopes in Antarctica, Chile, New Zealand and South Africa achieving \mbox{(sub-)milliarcsecond} (mas) angular resolution. 
The initial TANAMI sample was defined as a hybrid radio and $\gamma$-ray selected sample of AGN. Its main components are a radio selected flux-density limited subsample ($S_\mathrm{5\,GHz} > 2$\,Jy and $\alpha > - 0.5$ between 2.7\,GHz and 5\,GHz) and a $\gamma$-ray selected subsample of known and candidate sources based on results of CGRO/EGRET \citep{Hartman1999}. More detailed information about this project and its first epoch images and results is presented in \cite{Ojha2010a}.
Since the launch of \textit{Fermi}, new $\gamma$-ray detected, radio-loud AGN from the 1FGL-catalogue \citep{Abdo2010c} are continuously added to the sample which currently consists of 75 sources. 
% Currently, it consists of 75 sources including special sources of interest like the nearby radio galaxies Centaurus~A and Pictor~A, the VHE sources PKS\,2005$−$489 and PKS\,2155$-$304, or the gigahertz peaked spectrum (GPS) source PKS\,1718$-$649.

To achieve a simultaneous broadband coverage of the sample several additional monitoring programs at other wavelengths were set up to fill the gap between the simultaneous VLBI-\textit{Fermi}/LAT observations. At radio frequencies the monitoring with the Australia Telescope Compact Array (ATCA) provides flux density measurements and spectral information to e.g., determine the turnover frequency. In addition, with the Ceduna-Hobart Interferometer (CHI) we can quickly follow up  $\gamma$-ray flaring sources to measure simultaneous radio light curves \citep{Blanchard2012a}.
In the optical, the sample is part of the \textsl{Rapid Eye Mount} (REM, Istituto Nazionale di Astrofisica) program in order to provide spectra and redshifts. 

Altogether, the TANAMI program and its supporting multiwavelength campaigns will provide a unique broadband dataset of Southern Hemisphere AGN over the whole lifetime of \textit{Fermi} for extensive broadband emission and kinematic studies.

%%%%%%%%%%%%%%%%%%%%%%%%%%%%%%%%%%%%%%%%%%%%%%%%%%%%%%%%%%%
\section{TANAMI Dual-frequency VLBI-Observations}
\begin{figure*}
\includegraphics[width=0.33\textwidth]{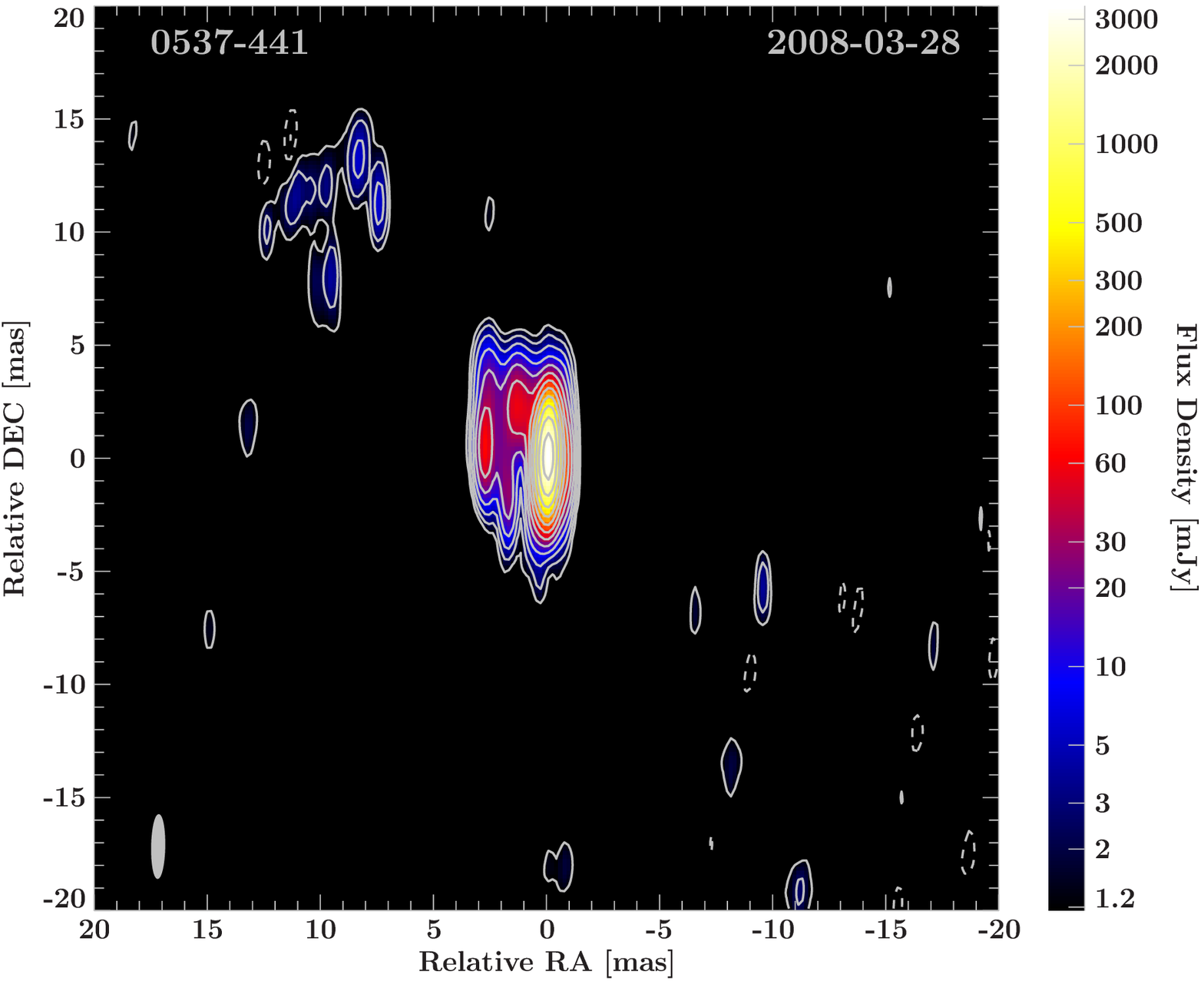}\hfill
\includegraphics[width=0.33\textwidth]{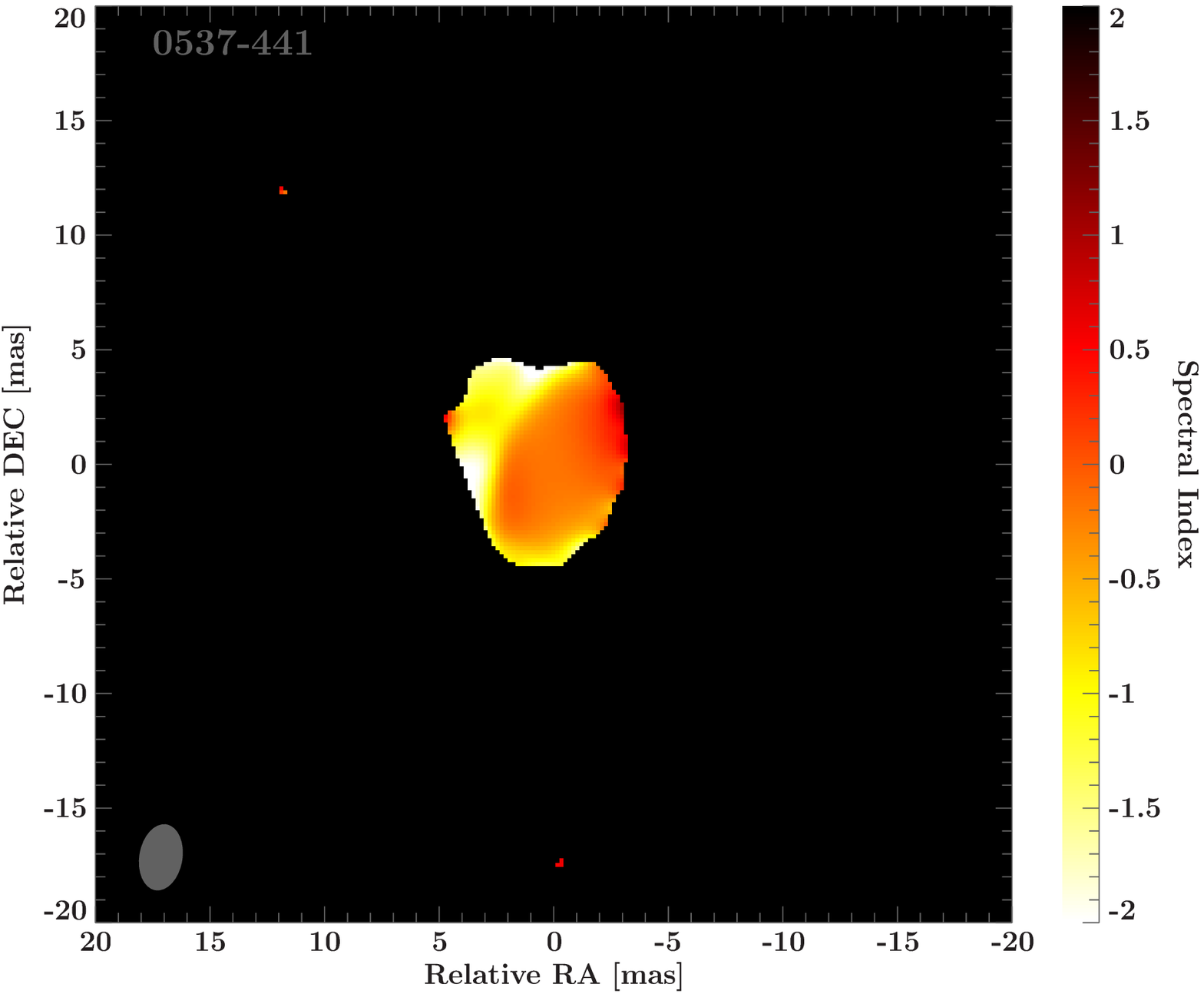}\hfill
\includegraphics[width=0.33\textwidth]{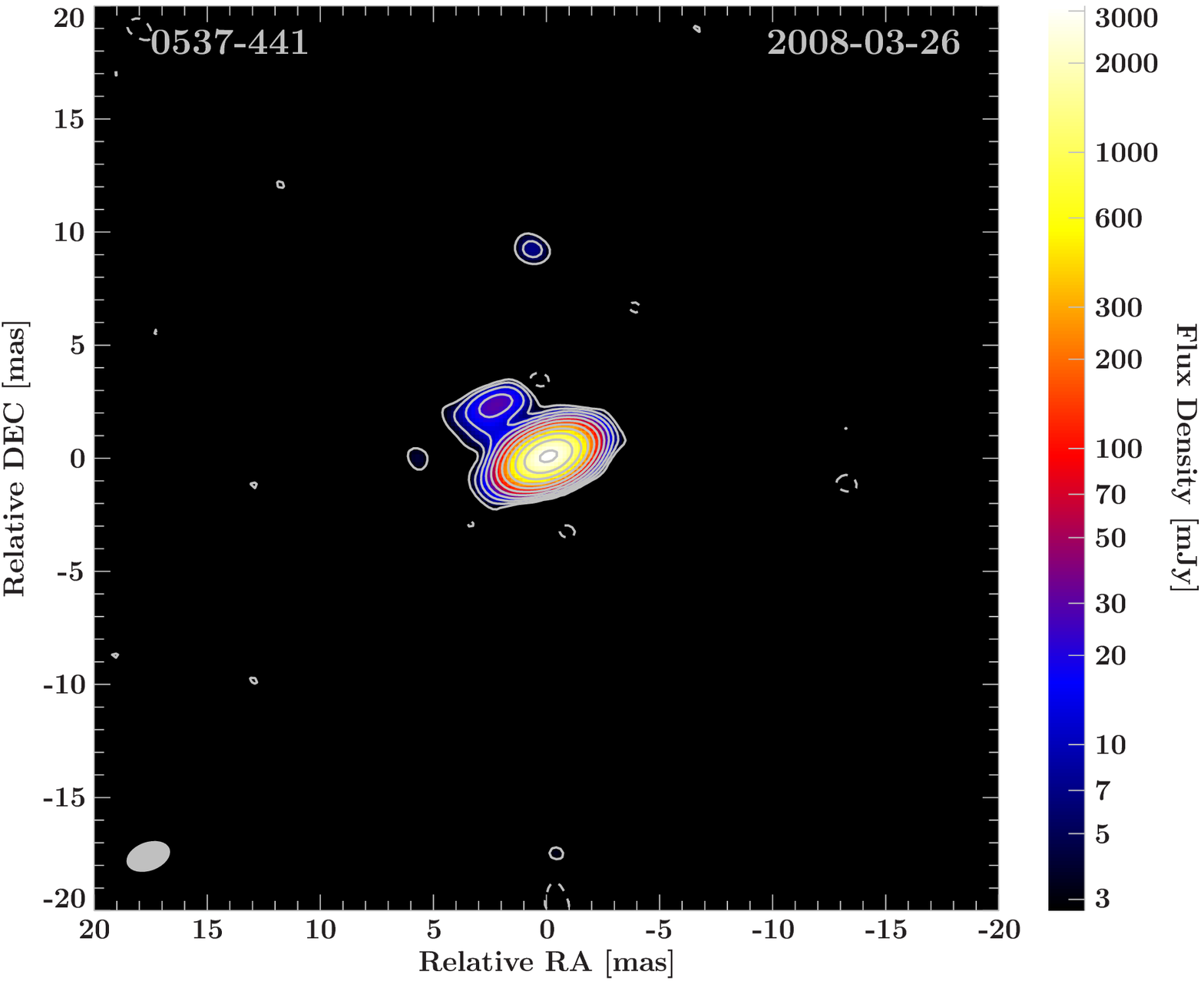}\hfill
\includegraphics[width=0.33\textwidth]{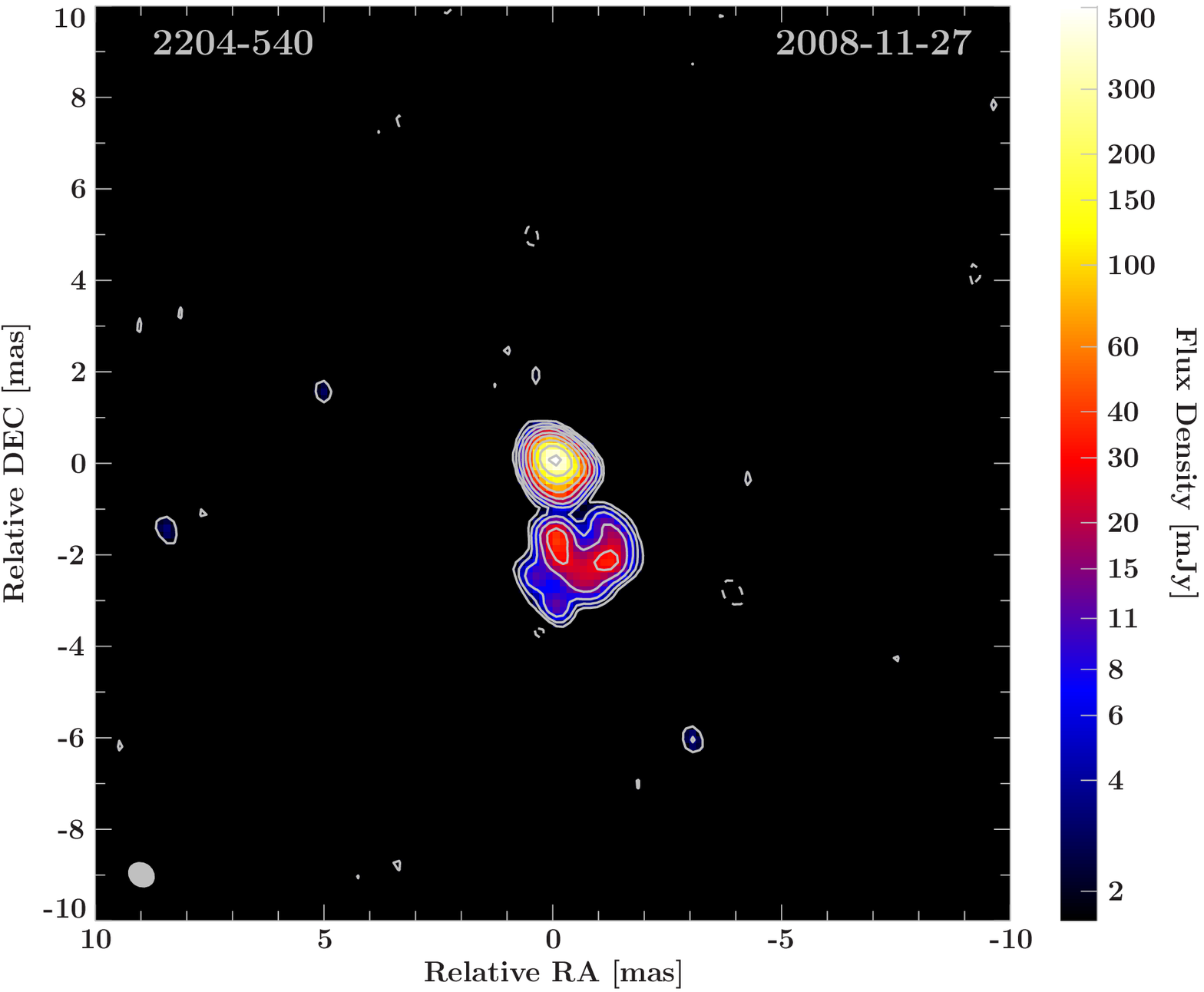}\hfill
\includegraphics[width=0.33\textwidth]{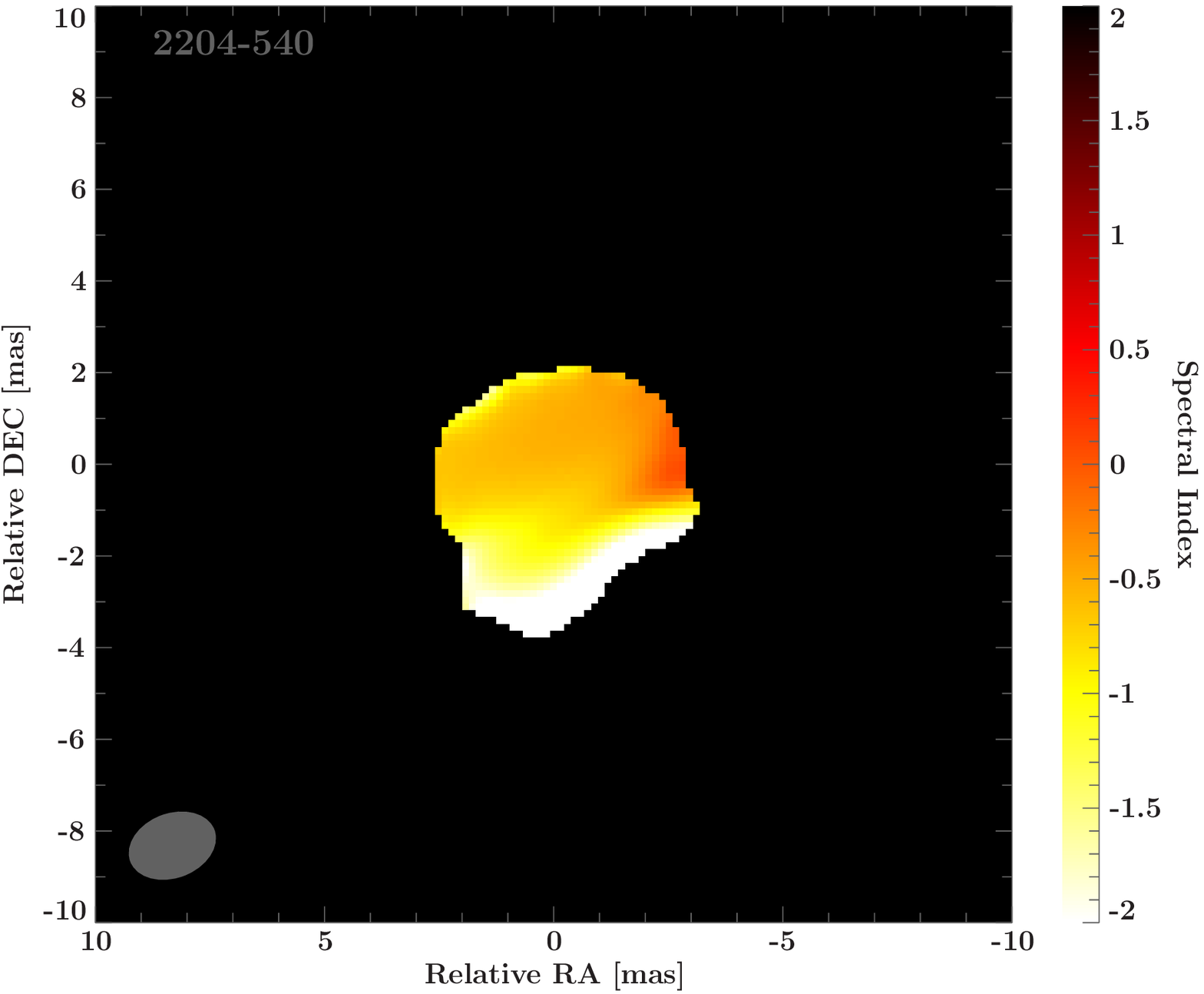}\hfill
\includegraphics[width=0.33\textwidth]{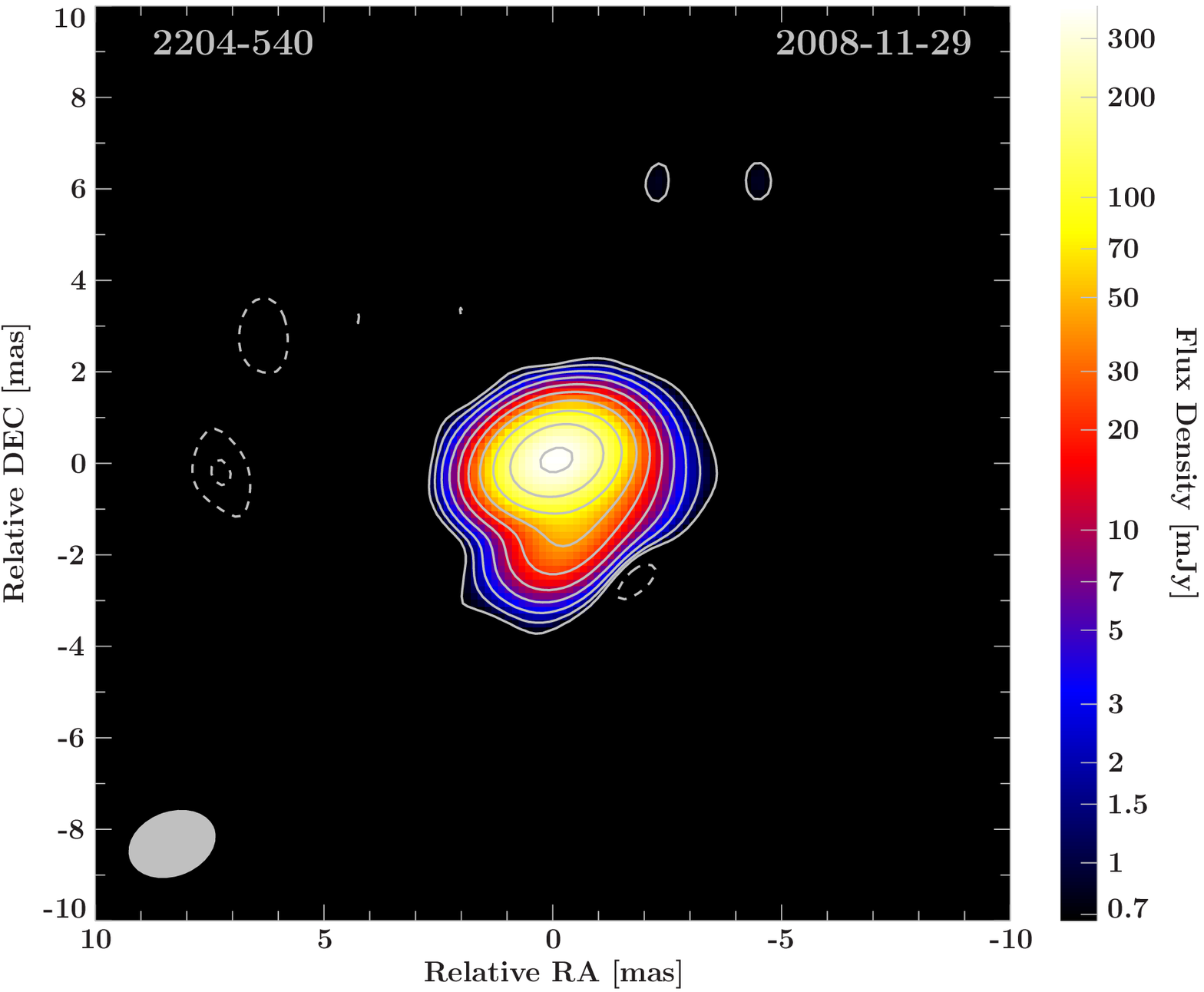}\hfill
\includegraphics[width=0.33\textwidth]{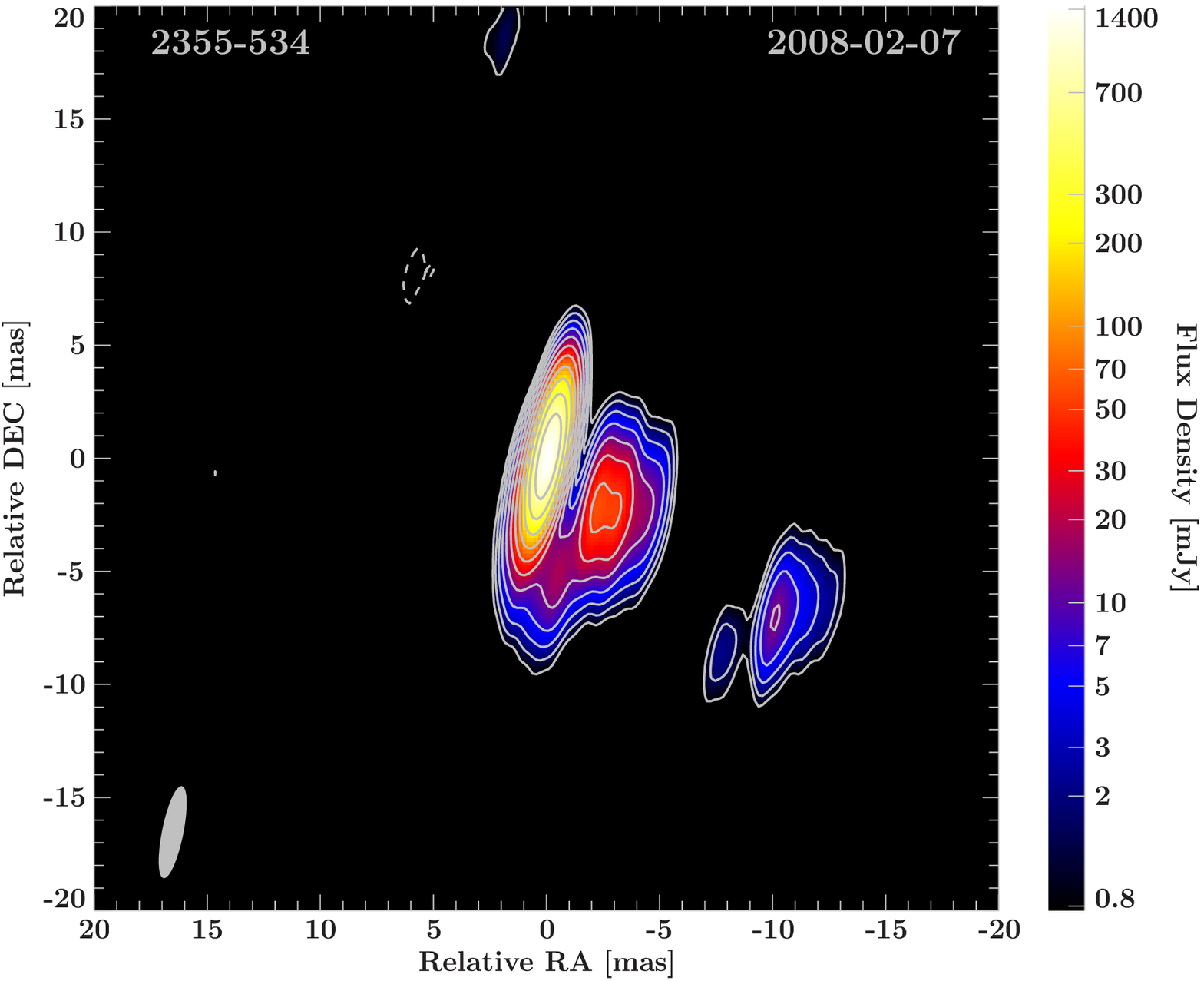}\hfill
\includegraphics[width=0.33\textwidth]{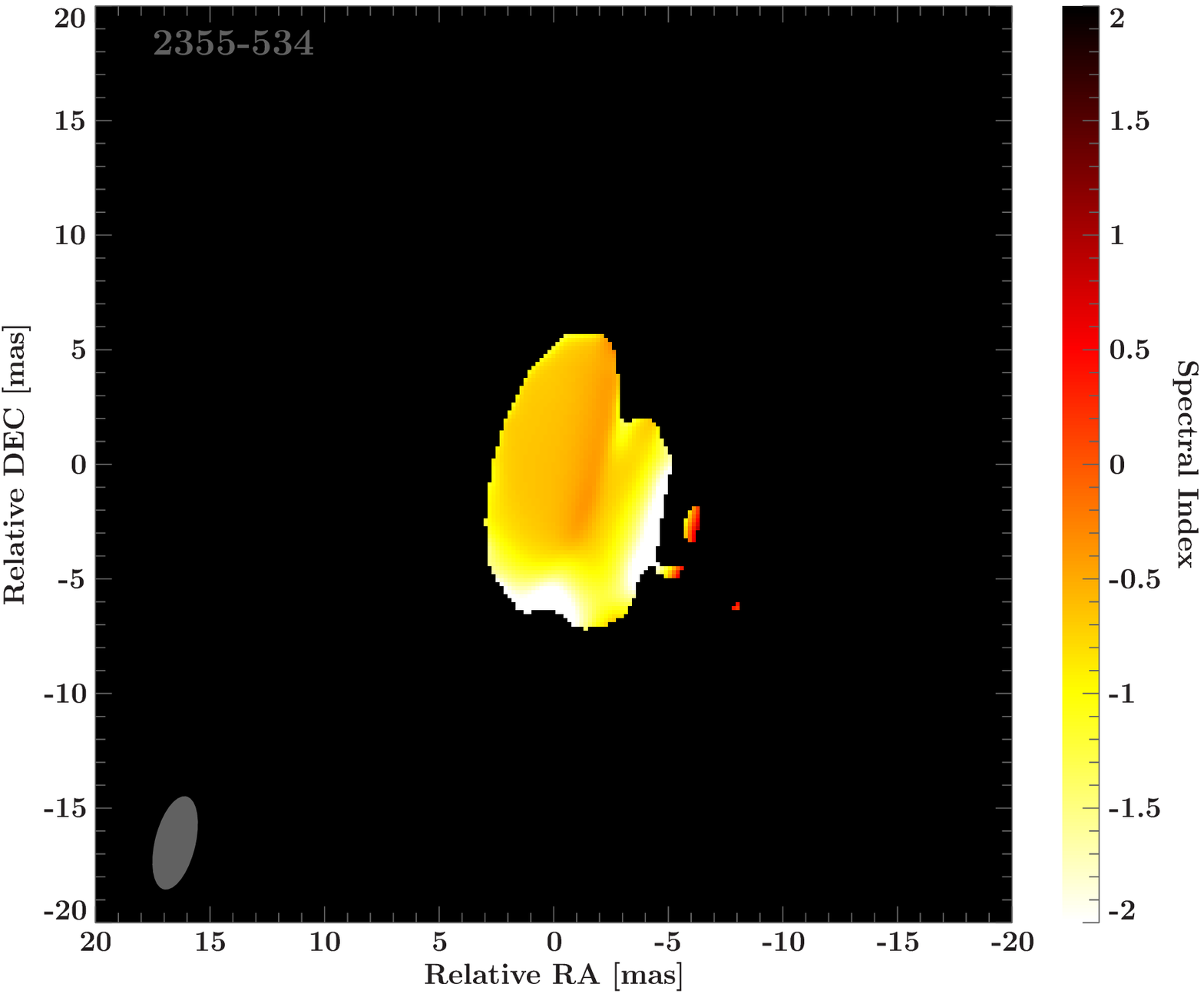}\hfill
\includegraphics[width=0.33\textwidth]{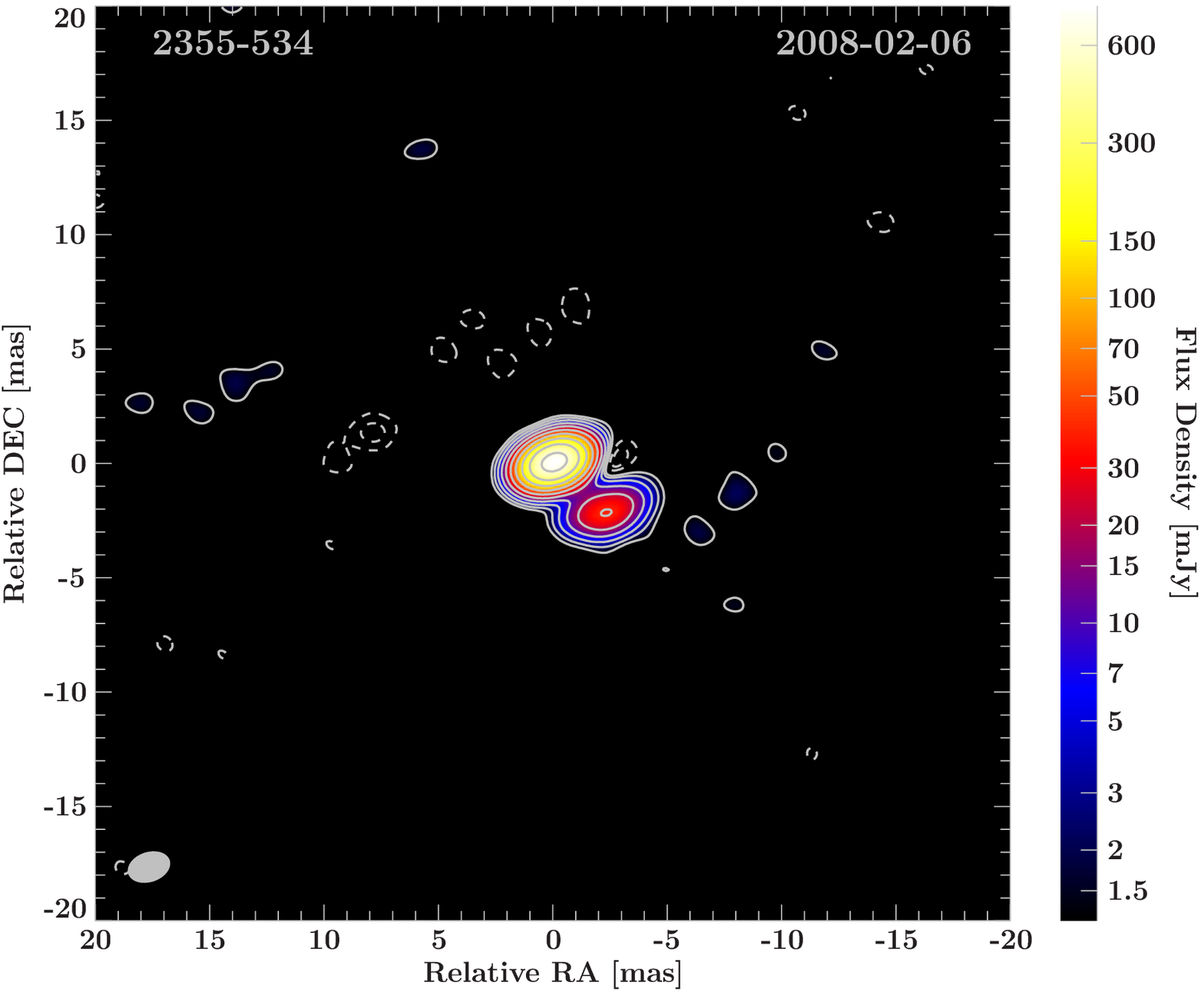}\hfill
\caption{Selection of simultaneous dual-frequency images and the corresponding spectral index maps (left: 8\,GHz image, middle: spectral index map, right: 22\,GHz image). Shown are the preliminary results for PKS\,0537$-$441 (top), PKS\,2204$-$540 (middle) and PKS\,2355$-$534 (bottom). Contours in the left and right panels are logarithmic, separated by a factor 2, with the lowest level set to the 3$\sigma$-noise-level (see also flux density scale bar to the right). The spectral index is derived where the flux densities at 8\,GHz and 22\,GHz exceed both the 3$\sigma$-noise-level. To construct the spectral index map both images have been restored with a common beam represented by the gray ellipse in the lower left corner. Note that no core-shift correction \citep{Lobanov1998} has been applied.}
\label{KXspix}
\end{figure*}

The special characteristic of TANAMI are the simultaneous dual-frequency observations at 8\,GHz and 22\,GHz of extragalactic jets south of $-30^\circ$ declination approximately every six months. This allows us to construct spectral index ($\alpha$ defined as \mbox{$F_\nu \propto \nu^{+\alpha}$}) maps with milliarcsecond resolution to study the spectral distribution and changes along the parsec-scale jets. Spectral index maps allow us to reveal optically thick ($\alpha \geq +0.7$) jet emission regions which are expected to be possible production sites for high energetic photons. Here we present some examples of the first dual-frequency images and the corresponding spectral index maps of the TANAMI sources (Fig.~\ref{KXspix}).
As expected the VLBI `core' of the sources has a flat or even inverted spectrum. 
The ongoing TANAMI monitoring will provide time dependent spectral index maps so that we can study the spectral index evolution of individual jet components with time providing further constraints on e.g.\ the correlation of high energy flares with spectral and structural changes within the jet.

%%%%%%%%%%%%%%%%%%%%%%%%%%%%%%%%%%%%%%%%%%%%%%%%%%%%%%%%%%%

\section{Sub-parsec Scale Kinematics of Centaurus A}
Centaurus~A is the closest AGN at a distance of $3.8 \pm 0.1$\,Mpc \citep{Harris2010} hosting a supermassive black hole of a mass of \mbox{$M=5.5\pm3.0\times 10^7 \mathrm{M}_{\odot}$} \citep{Israel1998, Neumayer2010}.  
Due to its proximity (1\,mas corresponds to just $\sim $$0.018$\,pc), Cen~A is exceptionally well suited for studying the innermost regions of AGN, the jet formation and collimation as well as testing different jet emission models. 

\begin{figure*}%[t]
\centering
\includegraphics[width=132mm]{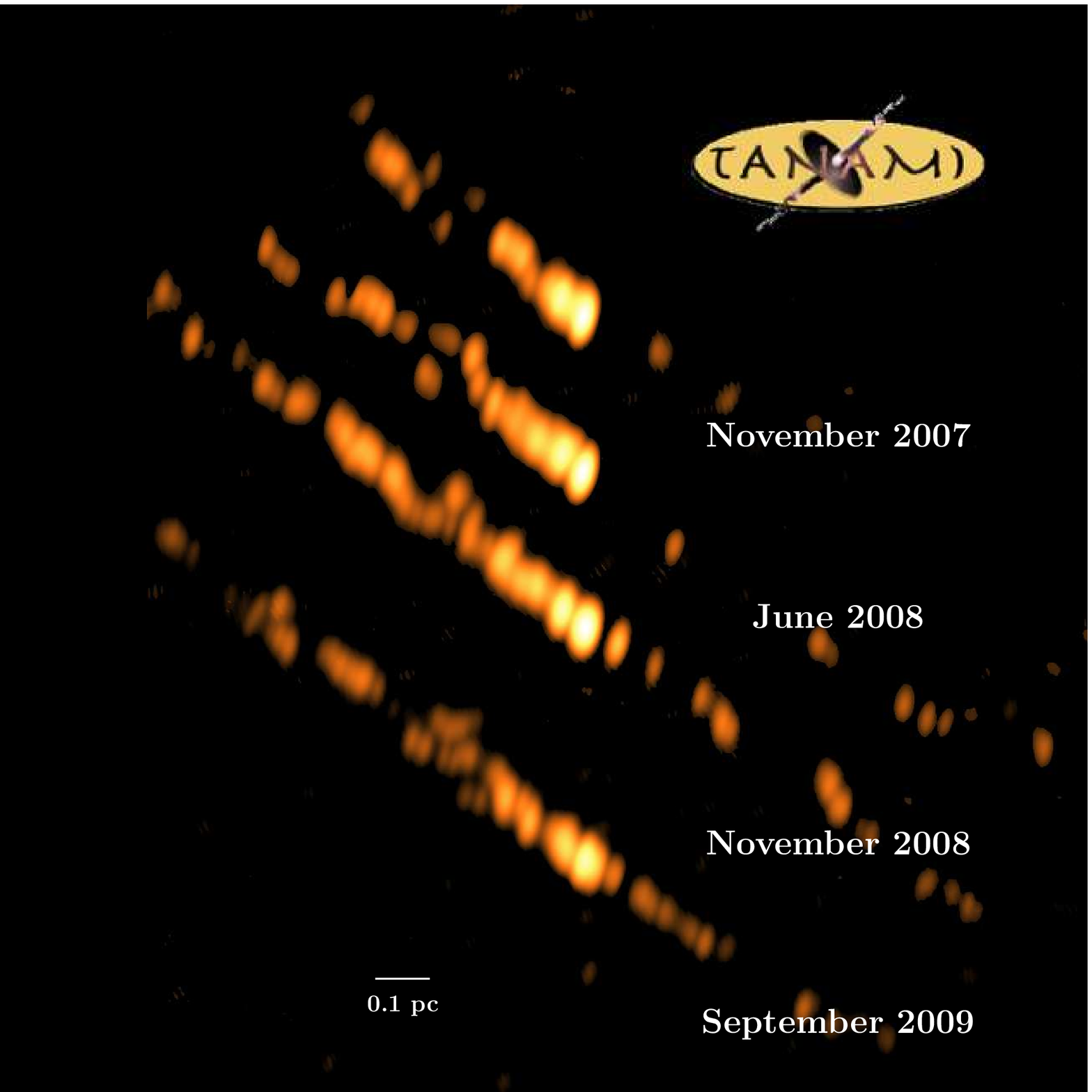}
\caption{Time evolution of Cen~A's jet and counterjet based on the first four 8\,GHz TANAMI images from 2007 November to 2009 September. All images \citep{Mueller2010, Mueller2011a} are restored with a common beam of ($2.86\times 1.16$)\,mas at a P.A.=$−13^\circ$. 
Displayed is emission above the 3$\sigma$-flux density level in each image. 
Two possible stationary features are detected: the second brightest feature next to the core at $\sim$$3.5$\,mas and the possible jet widening at $\sim$25\,mas.} \label{CenA_kinematics}
\end{figure*}

TANAMI has been monitoring Cen\,A at 8\,GHz twice a year since 2007 November including one simultaneous dual-frequency observations (2008 Nov.) with (sub-)milliarcsecond resolution. In Fig.~\ref{CenA_kinematics} we present the first four 8\,GHz observations showing the time evolution of individual jet components at sub-parsec scales.

With the first simultaneous dual-frequency TANAMI images at an angular resolution of about $0.7$\,mas$\times 0.4$\,mas (at 8\,GHz) we constructed the highest
resolution spectral index map of the jet-counterjet
system of Cen~A \citep{Mueller2011a}. We identify multiple possible emission sites within the inner few mas of the jet as the origin
of $\gamma$-ray photons from the Cen~A central region detected by \textit{Fermi}/LAT \citep{Abdocenacore2010}. This puts important constraints on SED models (single versus multi-zone models) for high energy emission in AGN jets.

We present first results of the multi-epoch analysis of Cen~A based on the first four high resolution TANAMI observations at 8\,GHz. By fitting Gaussian emission model components to the ($u,v$)-data, we can track individual, bright jet features of lightday-scale size revealing a very complex kinematical behaviour along the jet. 
The significant emission features within the sub-parsec scale jet seen in the 2007 November 8\,GHz TANAMI image are in good agreement with those in the images observed later on an approximately six months baseline \citep{Mueller2010, Ojha2010a}. The peak-flux densities at 8\,GHz indicate only a moderate radio flux variability: $\sim0.6$\,Jy in 2007 November, $\sim1.1$\,Jy in June 2008, $\sim0.7$\,Jy in 2008 November and $\sim0.5$\,Jy in 2009 September (cf.\ \cite{Mueller2010}).

We can track up to eight distinct, bright jet features. Our preliminary kinematical analysis gives a mean apparent jet speed of $\mu\approx 2.7\,\mathrm{mas/yr} \approx 0.16c$, that is consistent with previous results \citep{Tingay2001b}. 

In all images, we detect a prominent emission feature at $\sim$$3.5$\,mas downstream without significant proper motion. This might correspond to the stationary component detected by \citet{Tingay2001b} at $\sim$$4$\,mas. 
A possible widening and subsequent narrowing of the jet appears at $\sim$25\,mas ($\approx 0.45$\,pc) from the core in all 8\,GHz images, as well as in the 22\,GHz image \citep[cf.][]{Mueller2011a}. The 22\,GHz image suggests a co-spatial possible emission component. This feature might be interpreted as a stationary absorbing feature \citep[cf.][]{Mueller2011a}.
Noteworthy is also the overall similarity to the former space-VLBI image by \citet{Horiuchi2006} suggesting that at least some characteristic structures are moving slower than determined in earlier works \citep[$0.1c$--$0.45c$;][]{Tingay1998b, Tingay2001b, Hardcastle2003}, which would support the model of stationary features within the steady outflow.

Newer TANAMI observations are currently being analyzed to test and improve this first tentative model to describe the complex jet kinematics in detail \citep{Mueller2012prep}. With further dual-frequency observations we will also investigate the evolution of the spectral index along the jet and will combine spectral and structural changes with the $\gamma$-ray behaviour.
%%%%%%%%%%%%%%%%%%%%%%%%%%%%%%%%%%%%%%%%%%%%%%%%%%%%%%%
\begin{figure*}[ht]
\centering
\includegraphics[width=0.5\textwidth]{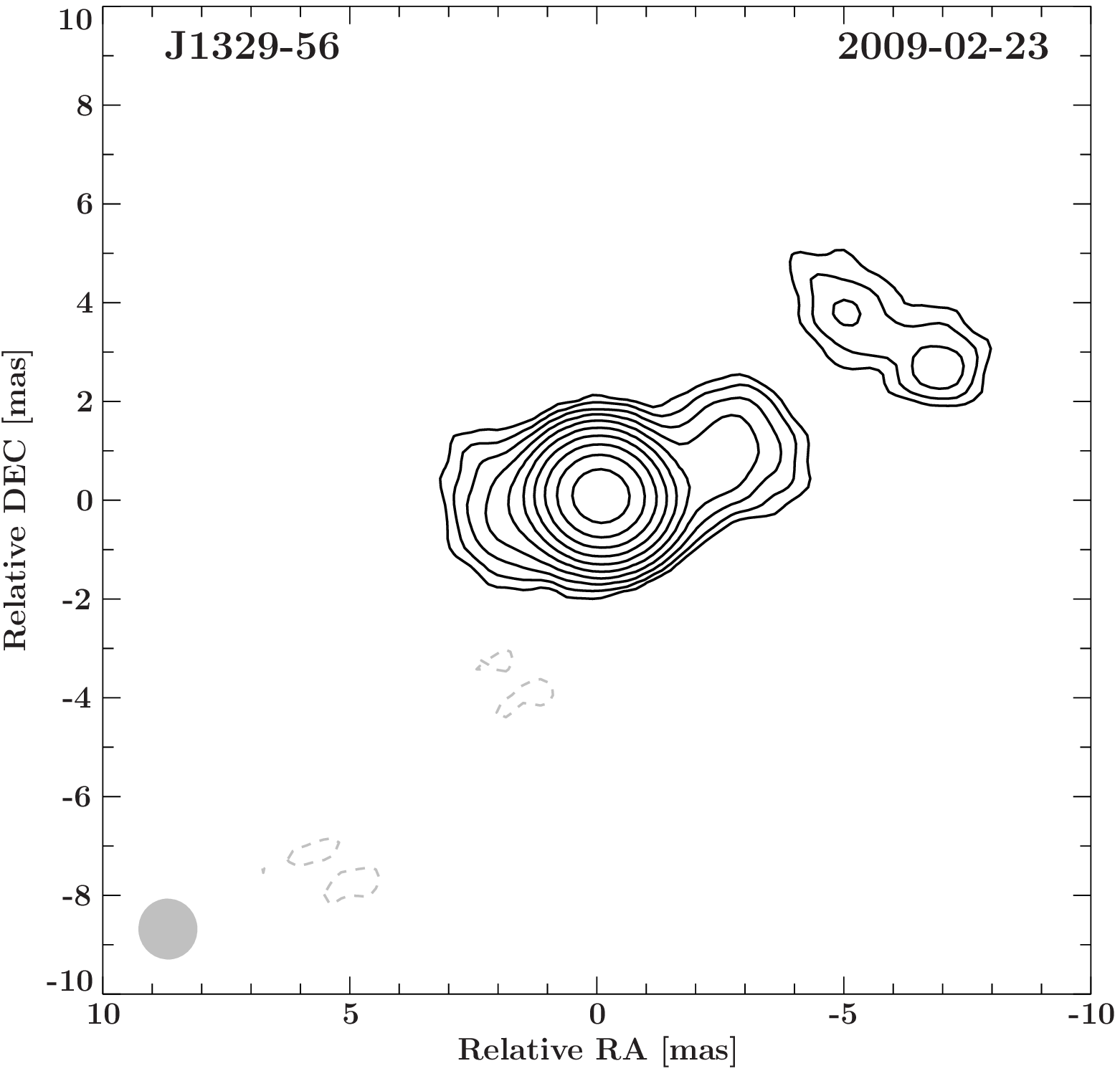}\hfill
\includegraphics[width=0.5\textwidth]{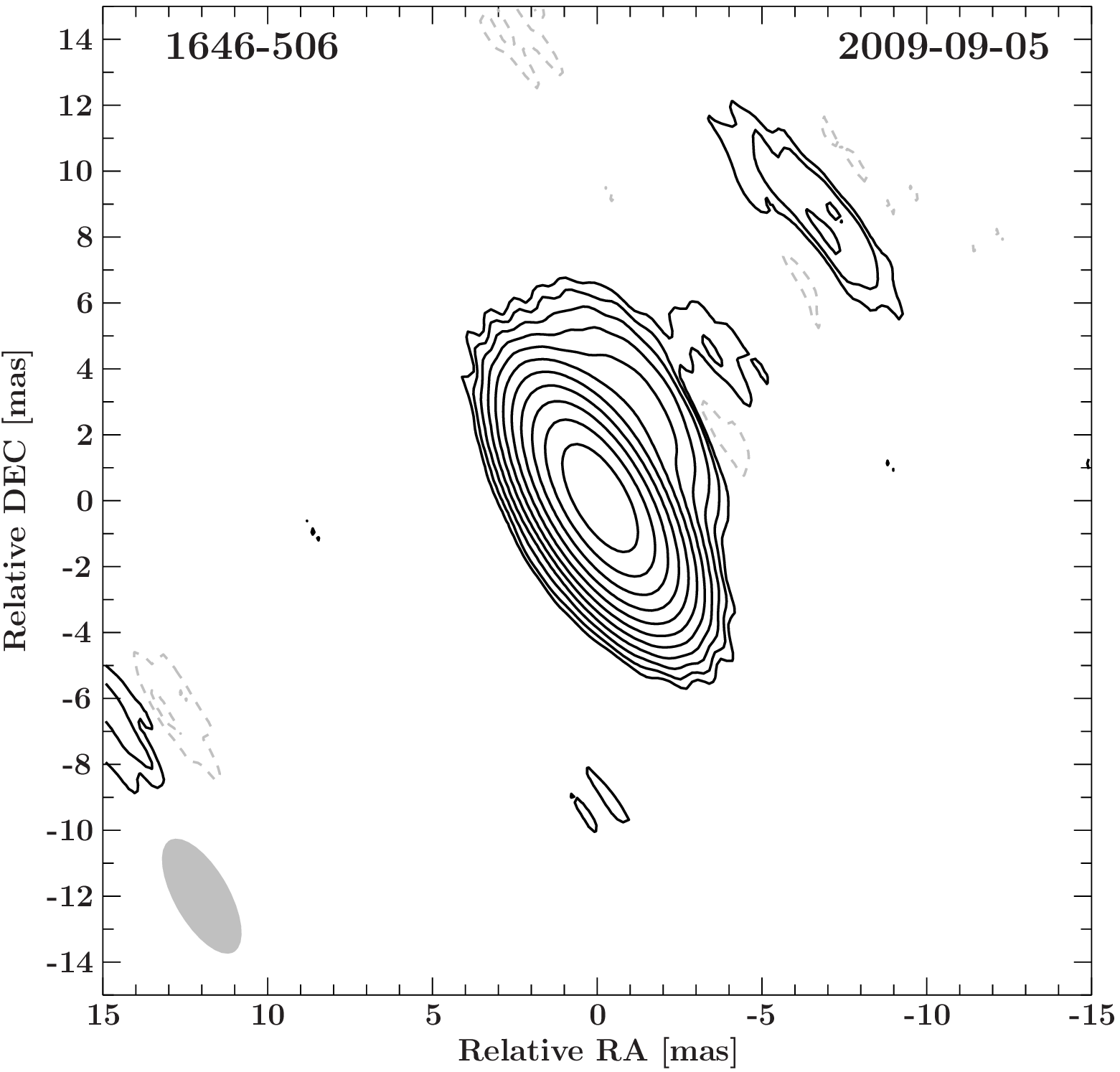}
\caption{TANAMI provides the first 8\,GHz VLBI images of new \textit{Fermi}/LAT-detected blazars south of $-30^\circ$ declination. As preliminary first results, we present here the mas-scale hybrid images of PMN~J1329$-$5608 ($S_\mathrm{peak}=0.19$\,Jy/beam) and PKS\,1646$-$558 ($S_\mathrm{peak}=1.0$\,Jy/beam) at 8\,GHz. The lowest contour defines the 3$\sigma$-flux density level. The contours are logarithmic with steps of factor 2.} \label{newsources}
\end{figure*}
\section{First VLBI-images of $\gamma$-bright sources newly detected by \textit{Fermi}/LAT}
Based on their $\gamma$-ray properties reported in the LBAS and the 1FGL-catalogue \citep{Abdo2009a, Abdo2010c}, 27 known radio-loud AGN associated with these new $\gamma$-ray sources were added to the TANAMI sample (see Table~\ref{table_newsources}) in order to investigate their radio structure with high resolution VLBI observations.
Since early 2009, TANAMI has been monitoring these sources about every three to six months, i.e.\ about four 8\,GHz epochs including one dual-frequency observations per source have already been performed. For most of these sources we provide the first VLBI images ever made. Here we present a selection of the first-epoch 8\,GHz images of these \textquoteleft new\textquoteright~sources (Fig.~\ref{newsources}). Further epochs are currently being calibrated and analyzed.

As being representatives of the $\gamma$-ray, bright TANAMI subsample, we present the first-ever VLBI images of PMN~J1329$-$5608 (PKS\,1325$-$558, 1FGL~J1329.2$-$5605) and PKS\,1646$-$558 (PMN~J1650$-$5044, 1FGL~J1650.4$-$5042) \citep{PMN1993, PMN1994, Fey2004, Massardi2008, Abdo2010c}. Both are optically unclassified blazar objects showing a bright compact core with a weak jet.

%%%%%%%%%%%%%%%%%%%%%%%%%%%%%%%%%%%%%%%%%%%%%%%%%%%%%%%%%%%%%%%%%
\section{Outlook}
TANAMI started in 2007 as the only large monitoring program of southern AGN supported by additional multiwavelength observations, primarily the $\gamma$-ray all-sky survey by \textit{Fermi}. Due to the contribution of the IVS antennas GARS (Antarctica) TIGO (Chile)\footnote{Operated by the German Bundesamt f\"ur Kartographie und Geod\"asie (BKG).}, and the LBA associated antennas at Tidbinbilla\footnote{Operated by the Deep Space Network of the National Aeronautics and Space Administration, USA.} the angular resolution of the TANAMI observations exceeds former Southern Hemisphere VLBI measurements. This enables us to produce the best resolved images of bright extragalactic jets south of $-30^\circ$ declination. Since 2011, the new Warkworth antenna\footnote{Operated by the Institute for Radio Astronomy and Space Research, Auckland University of Technology.} (New Zealand) considerably improves the ($u,v$)-coverage.  

Now, for almost the whole TANAMI sample the minimum required number of 8\,GHz observations are available to perform kinematic analysis. For the $\gamma$-ray bright subsample ejection times will be determined in order to investigate possible correlations with high energy flares. In addition, the first 22\,GHz epochs are currently being analyzed and simultaneous spectral index maps for all sources are being constructed. 

We started intense multiwavelength studies on individual sources, testing different proposed theoretical emission models for quasar SEDs. These try to explain the broadband emission seen from radio-loud AGN with leptonic or hadronic (or a combination of both) processes. Quasi-simultaneous multiwavelength monitoring of flaring and quiescence states of individual sources will allow us to experimentally determine which model is most likely to be correct (for more details, see contributions of Blanchard et al. \& Dutka et al. in these Conference Proceedings).
First TANAMI results were already contributed to several joint \textit{Fermi}-publications like e.g. \citet{Abdocenacore2010} or \citet{Abdo2009b}.

%%%%%%%%%%%%%%%%
\begin{table*}
\caption{Newly added TANAMI sources based on 1FGL} %% actually 31 added to 43 initial sample, but 27 are also in 1FGL
% \begin{tabular}{llccll}
\begin{tabular}{llr@{$\pm$}lr@{$\pm$}l}
 \hline
 \hline
%  Source  & 1FGL name & $S_\mathrm{1000}$$^\mathrm{a}$& $\Delta S_\mathrm{1000}$$^\mathrm{a}$& $\Gamma_\gamma$$^\mathrm{b}$ & $\Delta\Gamma_\gamma$$^\mathrm{b}$\\
Source & 1FGL name & \multicolumn{2}{c}{$S_{1000}^\mathrm{a}$} & \multicolumn{2}{c}{$\Gamma_\gamma$$^\mathrm{b}$} \\
 &  & \multicolumn{2}{c}{\tiny[$\times 10^{10}$ph/cm$^2$/s]} & \multicolumn{2}{c}{} \\
%    &  &  [$\times 10^{10}$ph/cm$^2$/s]$^\mathrm{a}$&  [$\times 10^{10}$ph/cm$^2$/s]$^\mathrm{a}$& & \\
 \hline
PKS 0055$-$328 & 1FGL J0058.4$-$3235 & 8 & 2 & 2.31 & 0.15 \\
 PKS 0227$-$369 & 1FGL J0229.3$-$3644 & 20 & 3 & 2.60 & 0.07 \\
 PKS 0244$-$470 & 1FGL J0245.9$-$4652 & 28 & 4 & 2.52 & 0.06 \\
 PKS 0302$-$623 & 1FGL J0303.4$-$6209 & 13 & 3 & 2.59 & 0.13 \\
 PKS 0308$-$611 & 1FGL J0310.1$-$6058 & 14 & 3 & 2.53 & 0.14 \\
 PMN J0334$-$3725 & 1FGL J0334.4$-$-3727 & 26 & 4 & 2.10 & 0.09 \\
 PKS 0402$-$362 & 1FGL J0403.9$-$3603 & 27 & 4 & 2.56 & 0.06 \\
 PMN J0413$-$5332 & 1FGL J0413.4$-$5334 & 24 & 4 & 2.55 & 0.08 \\
 PKS 0426$-$380 & 1FGL J0428.6$-$3756 & 257 & 10 & 2.13 & 0.02 \\
 PKS 0447$-$439 & 1FGL J0449.5$-$4350 & 111 & 7 & 1.95 & 0.04 \\
 PKS 0516$-$621 & 1FGL J0516.7$-$6207 & 32 & 4 & 2.28 & 0.09 \\
 PKS 0524$-$485 & 1FGL J0526.3$-$4829 & 15 & 3 & 2.37 & 0.11 \\
 PKS 0700$-$661 & 1FGL J0700.4$-$6611 & 47 & 5 & 2.15 & 0.07 \\
 PMN J0718$-$4319 & 1FGL J0718.7$-$4320 & 30 & 4 & 1.83 & 0.09 \\
 PKS 0736$-$770 & 1FGL J0734.1$-$7715 & 14 & 4 & 2.75 & 0.13 \\
 PMN J0810$-$7530 & 1FGL J0811.1$-$7527 & 25 & 4 & 1.80 & 0.11\\
 PKS 1057$-$79 & 1FGL J1058.1$-$8006 & 22 & 4 & 2.45 & 0.10 \\
 PKS 1101$-$536 & 1FGL J1103.9$-$5355 & 61 & 6 & 2.05 & 0.06 \\
 PMN J1329$-$5608 & 1FGL J1329.2$-$5605 & 41 & 6 & 2.56 & 0.07 \\
 PMN J1347$-$3750 & 1FGL J1347.8$-$3751 & 11 & 3 & 2.70 & 0.15 \\
 PKS 1440$-$389 & 1FGL J1444.0$-$3906 & 35 & 5 & 1.83 & 0.08 \\
 PMN J1604$-$4441 & 1FGL J1604.7$-$4443 & 77 & 8 & 2.46 & 0.05 \\
 PMN J1603$-$4904 & 1FGL J1603.8$-$4903 & 134& 11 & 2.12 & 0.04 \\
 PMN J1610$-$6649 & 1FGL J1610.6$-$6649 & 29 & 4 & 1.60 & 0.09 \\
 PMN J1617$-$5848 & 1FGL J1617.7$-$5843 & 23 & 5 & 2.72 & 0.10 \\
 PMN J1650$-$5044 & 1FGL J1650.4$-$5042 & 49 & 7 & 2.55 & 0.07 \\
 PMN J2139$-$4235 & 1FGL J2139.3$-$4235 & 83 & 6 & 2.08 & 0.04 \\
 \hline
\multicolumn{6}{l}{\footnotesize $^\mathrm{a}$ Photon flux and uncertainty for (1--100)\,GeV \cite{Abdo2010c}. } \\
\multicolumn{6}{l}{\footnotesize $^\mathrm{b}$ $\gamma$-ray spectral index and uncertainty \cite{Abdo2010c}.} \\
\end{tabular}
% \end{tabular}
\label{table_newsources}\\
\end{table*}

%%%%%%%%%%%%%%%%%%%%%%%%%%%%%%%%%%%%%%%%%%%%%%%%%%%%%%%%%%%%%%%%%%
\bigskip % extra skip inserted
\begin{acknowledgments}
% \small
This work has been partially supported by the Studienstiftung des
deutschen Volkes through a fellowship to C. M\"uller and by the European
Commission under contract ITN 215212 "Black Hole Universe". E.~Ros acknowledges partial support by the Spanish governement through grant AXA2009-13036-C02-02 and by the COST action MP0905 'Black holes in a violent Universe'. We thank the anonymous referee for thorough reading of the manuscript and useful comments.
\end{acknowledgments}

\bigskip % extra skip inserted
% % Create the reference section using BibTeX:
% %\bibliography{basename of .bib file}
\bibliographystyle{apsrev}
% \bibliography{cmbib,tanami,proc}

\end{document}